\documentclass[prl,twocolumn,floatfix,superscriptaddress]{revtex4}
\usepackage{dcolumn,amsmath}
\usepackage[dvips]{graphicx}
\usepackage{bm}
\usepackage{hyperref}  

\newcommand{\dso}[0]{$\Delta_{SO}$}
\newcommand{\dbp}[0]{$\Delta_{SO}^{\mathrm{B88P86}}$}
\newcommand{\dpb}[0]{$\Delta_{SO}^{\mathrm{PBE0}}$}

\begin{document}

\title{Interaction of copernicium with gold:
        Assessment of applicability of simple density functional theories}

\author{Andr\'ei Zaitsevskii}
\affiliation{FSBI ``Petersburg Nuclear Physics
             Institute'', Gatchina, Leningrad district 188300, Russia}
\affiliation{NRC ``Kurchatov Institute'', Kurchatova sq. 1, Moscow,
             123182 Russia}
\author{Anatoly V. Titov} 
\affiliation{FSBI ``Petersburg Nuclear Physics Institute'', Gatchina, Leningrad
             district 188300, Russia}

\date{\today}

\begin{abstract}
Interactions of Cn (element 112) atoms with small Au clusters are studied using
accurate \emph{ab initio} scalar relativistic coupled cluster method for
correlation treatment and two-component relativistic density functional theory
(RDFT) to account for spin-dependent relativistic effect. The results
demonstrate the failure of RDFT with simple generalized-gradient and hybrid
functionals in describing Cn--Au bonds in complex systems.
\end{abstract}

\maketitle

Successful identification of superheavy element 112 (Cn) by thermochromatography
on gold surface \cite{eichler:07,eichler:08} had resulted in an avalanche of
theoretical studies of the nature and properties of the interactions of the Cn
atom with gold (see \cite{pershina:09,pershina:11,
zaitsevskii:10} and references therein). It is generally believed (see, e.g.,
\cite{pershina:11}) that the relativistic density functional theory (RDFT) with
simple generalized gradient approximations (GGA) for the exchange-correlation
functional (XCF) \cite{becke:88,perdew:86} provides a satisfactory tool for
theoretical modeling of Cn/Au surface complexes.  Since the quantitative
experimental information on Cn\,--\,Au interactions is restricted to a single
measured value, the adsorption temperature of Cn on gold surface, thus being
obviously insufficient to estimate the reliability of theoretical approaches,
the main argumentation for the validity of RDFT/GGA in applications to their
description relies on the data concerning diatomic systems (see
e.g.~\cite{zaitsevskii:06}) and on the experience of calculations of other heavy
element compounds.  It should be noticed, however, that the bulk of such
experience is nearly useless because of unique features of the Cn atomic
structure in the valence region.  The presence of relatively compact filled
$d$-shell with rather high orbital energies (resembling that in Au and Hg)
enables one to expect unusually strong $d_{\rm Cn}^{10}-d_{\rm Au}^{10}$
contributions to the Cn--Au bonding, in a sense similar to the aurophilic
attraction \cite{schwarz:04}.  The efficiency of the RDFT with semilocal
functionals in such cases is at least questionable
as demonstrates the example of the Au$_n$ cohesion energy which is nicely
reproduced by RDFT/GGA in the Au diatomic but becomes progressively
underestimated for larger clusters and bulk \cite{rusakov:07,haeberlen}.  The
dependence of the accuracy of RDFT/GGA based results for gold clusters on the
coordination numbers of Au atoms has been shown in Ref.~\cite{rusakov:07}. 

The mentioned facts inspire serious doubts concerning the adequacy of the
RDFT/GGA treatment of the systems under discussions. These doubts are
strengthened by the detailed  pilot comparison of the results of relativistic
DFT and many-body perturbation theory (MBPT) calculations on moderate-size
Hg--Au$_n$ and Cn--Au$_n$ complexes \cite{zaitsevskii:10}.  Unfortunately,
finite-order MBPT binding energies for similar (Au- and Hg-containing) systems
are strongly affected by convergency problems and the resummation scheme chosen
in this work was not optimal \cite{pade}. The errors due to the use of rather
restricted one-electron basis sets with insufficient flexibility of high angular
momentum components can also be non-negligible.  A more accurate correlation
treatment is, therefore, required to pronounce the decisive verdict.

In the present Letter we report the calculations on Cn-Au$_n$ complexes,
$n=1{-}4$, by a hybrid scheme combining \emph{ab initio} scalar relativistic
correlation calculations with the estimation of spin-dependent relativistic
effects (effective spin-orbit couplings) through geometry-dependent corrections
to interaction energies ($\Delta_{SO}$) obtained at the RDFT level
\cite{zaitsevskii:06}.  The necessary accuracy of the correlation treatment is
ensured by the use of an accurate coupled cluster technique and the
extrapolation to the complete basis set limit. We have to stress that our scheme
does not imply the smallness of spin-dependent interactions or the neglect of
their interference with electron correlations \cite{russchemrev:08}.  The
geometries of polyatomic systems ($n=2{-}4$) were restricted to those resembling
possible structures of adsorption complexes on the gold (111) surface, namely,
the $C_{2v}$ configurations of Cn--Au$_2$ and Cn--Au$_4$, as well as the
$C_{3v}$ configuration of Cn--Au$_3$ simulating the ``bridge'' and ``hollow''
positions of the Cn adatom, respectively. The Au--Au distances were fixed at
their bulk value (2.884~\AA); only the distances $r$ between the Cn atom and the
center of the Au subsystem were optimized.


The computational scheme employed in the present study resembles that used in
Ref.~\cite{zaitsevskii:11}.  The ``small'' atomic cores (60 electrons for Au and
92 electron for Cn) were replaced by accurate shape-consistent semilocal
relativistic pseudopotentials (RPPs) \cite{mosyagin:06amin,mosyagin:10}
optimized for the description of valence shells.  Scalar relativistic
calculations were performed with the spin-averaged version of these RPPs by the
conventional coupled cluster method with fully optimized single and double and
perturbative triple amplitudes, CCSD(T), implemented in the {\sc cfour} program
package \cite{cfour:10}.  For the systems with odd numbers of gold atoms
unrestricted Hartree--Fock references were used.  The innermost explicitly
treated atomic shells ($6s6p$ for Cn and $5s5p$ for Au) were not correlated. 

We used the correlation-consistent basis sets (\cite{zaitsevskii:11,bases:11})
constructed according to the prescriptions from Ref.~\cite{puzza:05}. Taking
into account the experience of calculations on Au-Au and Au-Hg
\cite{puzza:05,pade}, the bases were augmented by diffuse functions (the
``aug-cc'' variant).  The complete basis set limit (CBS) estimates were derived
from the energies obtained with triple and quadruple zeta quality bases using
the conventional two-point formula \cite{cbs:97}. As has been pointed out in
Ref.~\cite{zaitsevskii:11}, the extrapolation procedure efficiently suppresses
basis set superposition errors, so that we had no need to invoke any
counterpoise procedure.

The contributions from the spin-dependent relativistic effects to the
interaction energies were evaluated by comparing the results of one- and
two-component (1c and 2c) non-collinear RDFT calculations performed with the
code \cite{wuellen:10}. The data presented in this letter were obtained with
simple generalized gradient approximation \cite{becke:88,perdew:86} usually
denoted by the acronym B88P86 and believed to be well suited for superheavy
element electronic structure calculations, and with PBE0 hybrid functional
\cite{adamo:99} chosen because of its (partially) non-empirical nature. 

The main results of the calculations are listed in Table~\ref{maintable}. The
equilibrium separations and binding energies derived from 1c CCSD(T)/CBS+\dso{}
potential curves for the CnAu diatomic molecule are in very reasonable agreement
with those obtained by RDFT with both XCFs employed. When the cluster size
increases, the 2c RDFT binding energies deviate \emph{progressively} from their
1c CCSD(T)+\dso{} counterparts; for the largest system under study the
underestimation of the binding energy reaches ca.\ 40-45~\%.  It might be
interesting to notice that the results of a similar numerical experiment with Cn
replaced by the neighbor element 113 (E113) with markedly lower energies of the
filled $d$ shell do not exhibit a similar trend.

\begin{table}[h!]
\caption{Equilibrium $r$ values  (\AA) and Cn-Au$_n$ binding energies (eV).}
\begin{center}
\begin{tabular}{lcccccc}
                           &    CnAu         &  CnAu$_2$ & CnAu$_3$ & CnAu$_4$ \\
                           &                    &                    \\
$r_e$, \hspace{0.35em}2c RDFT/B88P86  &2.73 &       2.65             & 2.61 &   2.68     \\
\hspace{1.85em}2c RDFT/PBE0    &   2.74    & 2.72    & 2.63 &  2.68   \\
\hspace{1.85em}1c CCSD(T)+\dbp{} &  2.73   &     2.63 & 2.59  &   2.64         \\
\hspace{1.85em}1c CCSD(T)+\dpb{} &  2.72    &     2.65 & 2.60 &  2.67     \\
\\
\hspace{1.85em}MBPT+\dbp{}$\,^{*)}$ &2.78 &&&2.60 \\
\hspace{1.85em}MBPT+\dpb{}$\,^{*)}$ &2.77 &&&2.61 \\
                           &                    &                    \\
$D_e$, 2c RDFT/B88P86             &  0.47  &  0.23  & 0.29 &  0.25 \\
\hspace{1.85em}2c RDFT/PBE0    &   0.39       & 0.19  & 0.28 &  0.26 \\
\hspace{1.85em}1c CCSD(T)+\dbp{} &   0.45    & 0.33  & 0.45  &  0.46\\
\hspace{1.85em}1c CCSD(T)+\dpb{} &    0.42     & 0.29  & 0.42 & 0.42 \\
\\
\hspace{1.85em}1c MBPT+\dbp{}$\,^{*)}$ &0.37&&& 0.60 \\
\hspace{1.85em}1c MBPT+\dpb{}$\,^{*)}$ &0.35&&& 0.55 \\
\end{tabular}
\end{center}
$^{*)}$ Ref.~\cite{zaitsevskii:10}; fourth-order MBPT was followed by the extrapolation
using constrained [3/2] Pad\'e approximant.  
\label{maintable}
\end{table}

\begin{figure}[h!]
\begin{center}
\includegraphics[width=0.47\textwidth,clip=true]{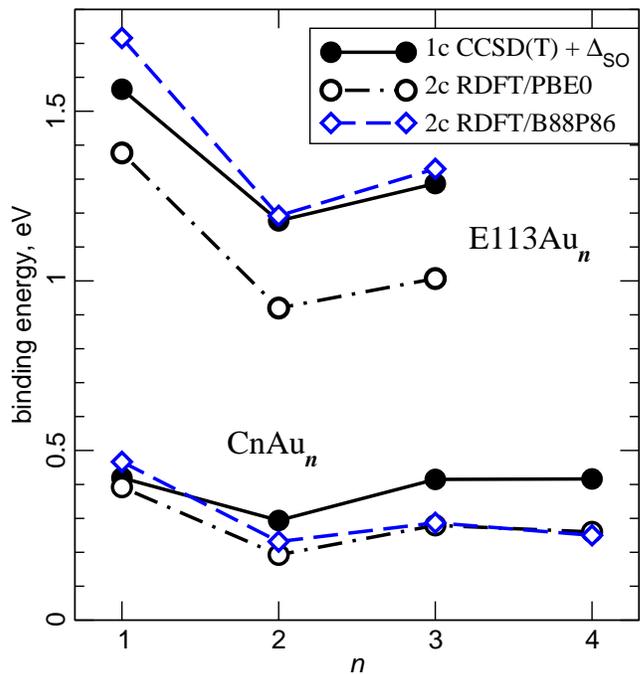}
\caption{The dependencies of CCSD(T)+$\Delta_{SO}$ and RDFT binding energies on gold cluster size.
The data on E113--Au$_n$ systems are taken from Ref.~\cite{zaitsevskii:11}.}
\label{versus}
\end{center}
\end{figure}

The extrapolated MBPT+\dso{} binding energies for Cn--Au$_4$ from
Ref.~\cite{zaitsevskii:10} are visibly too large; nevertheless, the main
conclusion of the cited work concerning the failure of RDFT with simple GGA and
GGA-based hybrid energy functionals in describing more or less complex Cn--Au
systems is confirmed by the present accurate calculations. A reasonable
agreement between the most recent RDFT/GGA Cn/Au adsorption energy estimate
(0.46 eV, \cite{pershina:09}) and the presently assumed experiment-based value
(0.54$^{+4}_{-3}$ eV, \cite{eichler:08}) seems not to be sufficient to prove the
opposite opinion and, probably, results from a fortuitous cancellation of errors
in \cite{pershina:09} (for instance, those of approximate nature of the B88P86
functional and of use of purely local functional to get the electron and
magnetization density distributions).    

\medskip\noindent
The work was supported by the Russian Foundation for Basic Research (Grant No.\
11-03-12155-ofi-m-2011).

\end{document}